\documentclass[10pt,journal,cspaper,compsoc]{IEEEtran}

\ifCLASSINFOpdf
  \usepackage[pdftex]{graphicx}
\else
\fi

\usepackage{amssymb,amsmath}
\usepackage{color}
\usepackage[colorlinks=true]{hyperref}

\newcommand{\ri}{{\em i)} }
\newcommand{\rii}{{\em ii)} }
\newcommand{\riii}{{\em iii)} }
\newcommand{\riv}{{\em iv)} }
\newcommand{\rv}{{\em v)} }

\newcommand{\rx}{$\times$ }
\newcommand{\eg}{{\em e.g.}, }
\newcommand{\ie}{{\em i.e.}, }

\usepackage{multirow}
\usepackage{tikz}
\usetikzlibrary{calc}

\begin{document}
% TITLE
\title{Graph Representation for Face Analysis in Image Collections}

% AUTHORS
\author{Domingo~Mery
        and~Florencia~Valdes% <-this % stops a space
\IEEEcompsocitemizethanks{\IEEEcompsocthanksitem D.Mery and F.Valdes are with the Department
of Computer Science, Pontificia Universidad Cat\'{o}lica, Santiago of Chile.\protect\\
URL: http://domingomery.ing.uc.cl}
%\IEEEcompsocthanksitem K.Bowyer is with Department of Computer Science \& Engineering, University of Notre Dame.}% <-this % stops a space
\thanks{}}

%\markboth{IEEE Trans. on Pattern Analysis and Machine Intelligence}%
%{Shell \MakeLowercase{\textit{et al.}}: Bare Demo of IEEEtran.cls for Computer Society Journals}

% ABSTRACT
\IEEEcompsoctitleabstractindextext{

\begin{abstract}

Given an image collection of a social event with a huge number of pictures, it is very useful to have tools that can be used to analyze how the individuals --that are present in the collection-- interact with each other. In this paper, we propose an optimal graph representation that is based on the `connectivity' of them. The connectivity of a pair of subjects gives a score that represents how `connected' they are. It is estimated based on co-occurrence, closeness, facial expressions, and the orientation of the head when they are looking to each other. In our proposed graph, the nodes represent the subjects of the collection, and the edges correspond to their connectivities. The location of the nodes is estimated according to their connectivity (the closer the nodes, the more connected are the subjects). Finally, we developed a graphical user interface in which we can click onto the nodes (or the edges) to display the corresponding images of the collection in which the subject of the nodes (or the connected subjects) are present. We present relevant results by analyzing a wedding celebration, a sitcom video, a volleyball game and images extracted from Twitter given a hashtag. We believe that this tool can be very helpful to detect the existing social relations in an image collection.

\end{abstract}

}

\maketitle
\IEEEdisplaynotcompsoctitleabstractindextext
\IEEEpeerreviewmaketitle

\section{Introduction}
\label{Sec:Introduction}

% In face recognition, the task is to identify a subject appearing in an image as a unique individual. Over the last decade, we have witnessed tremendous improvements in face recognition algorithms. Some applications, that might have been considered science fiction in the past, have become reality now. However, it is clear that face recognition, is far from perfect when tackling more challenging images such as faces taken in unconstrained environments \eg face images acquired by long-distance cameras. Although innovative methods in computer vision have improved the state of the art, the performance obtained in low-quality images remains unsatisfactory for many applications. 

\begin{figure*}[t!]
\centering 
\includegraphics[width=0.95\textwidth]{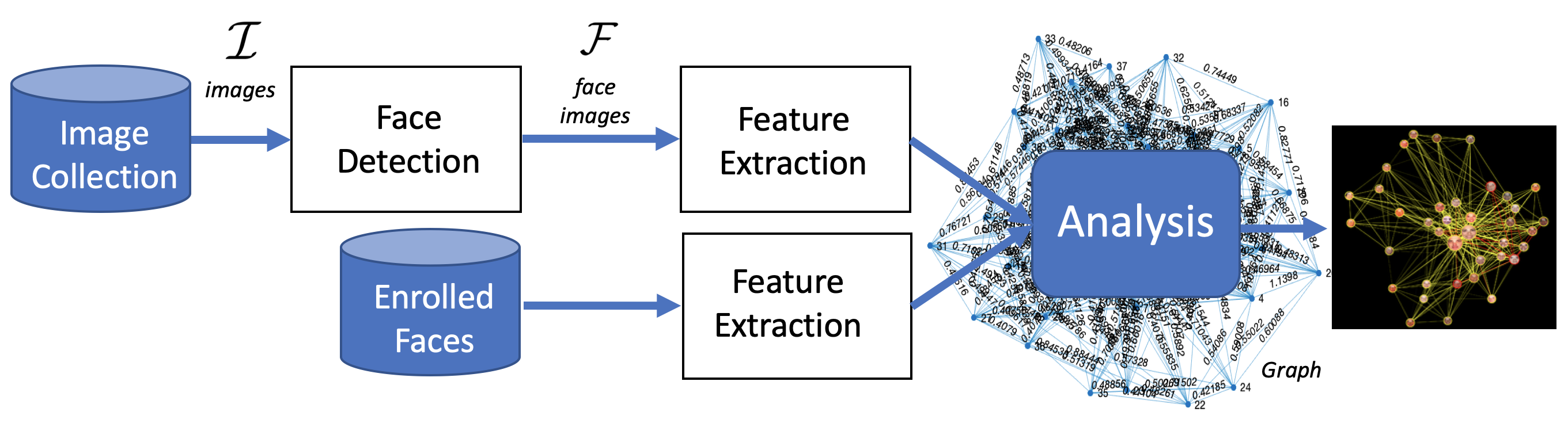}
\caption{Block diagram of our proposed method.}
\label{Fig:BlockDiagram}
\end{figure*}

In the last decades, it has been a common practice to have collections of a huge number of images \cite{Mery2019:PSIVT}. For example, a social event --like a wedding celebration-- can have thousands of pictures. Moreover, we can download thousands of images from social networks given a hashtag. %, a keyword or a location. 
In practice, it is not so simple to browse all images, and it can be very helpful to analyze the faces and build a graph representation with information extracted from all faces that are present in the images as illustrated in Fig. \ref{Fig:BlockDiagram}, where the nodes represent the subjects and the edges are the connections between them.% Thus, we can build a relational representation based on the face attributes and connection information extracted from them. 

In recent years, we have witnessed tremendous improvements in face recognition by using complex deep neural network architectures trained with millions of face images (see for example advances in face recognition \cite{parkhi_2015} \cite{deng2019arcface} \cite{li2019low} and in face detection \cite{zhang2016joint} \cite{sun2018face}). In addition, there are very impressive advances in applications \cite{mery2019student}, face clustering \cite{otto2017clustering} \cite{shi2018face}, and in the recognition of age \cite{rothe2018deep}, gender \cite{zavan2019benchmarking}, facial expressions  \cite{arriaga2017real}, eye gaze \cite{visapp19}, head pose \cite{Ruiz_2018_CVPR_Workshops} and facial landmarks \cite{zhu2019robust}.

In this field, many works deal with applications that can be developed using the face analysis tools. Here some examples:
In \cite{renoust2016face}, social networks are built by detecting and tracking faces in news videos, the idea is to establish how much and with whom a politician appears on TV news.
In \cite{baltrusaitis2018openface}, a facial behavior analysis is presented. The method can extract expressions and action units (facial movements such as `inner portion of the brows is raised' or `lips are relaxed and closed'), that could be used to build interactive applications. 
In \cite{zhang2015learning}, `social relations' (defined as the association like warm, friendliness and dominance, between two or more persons) are detected in face images in the wild. In \cite{huang2018multimodal}, face-based group-level emotion recognition is proposed to study the behavior of people participating in social events. Similar advances have been made in video analysis using the information of the location and actions of people in videos. See for example \cite{hoai2014talking}, where a `context aware' configuration model is proposed for detecting groups of persons and their interactions (\eg handshake, hug, etc.) in TV material. 

It is evident that the state of the art in this field is far from perfect, and any contribution that can help to detect and measure the interaction of individuals in an image collection will play a relevant role when analyzing social relations. 

Our contributions are threefold. Given an image collection and a set of subjects that are present in the collection: \ri We define a `connectivity' measurement between every pair of subjects of the set that gives a score that represent how connected they are. The connectivity is based on co-occurrence, closeness, facial expressions, and the orientation of the head when they are looking to each other. \rii We estimate an optimal graph in which the nodes represent the subjects of the set, and the length of the edges that connect the nodes depends on the connectivity of the subjects (the closer the nodes, the more connected are the subjects). \riii We developed a graphical user interface in which by clicking onto the nodes (or the edges), the user can display the corresponding images of the collection in which the subject of the nodes (or the connected subjects) are present.

\section{Proposed Method}
\label{Sec:ProposedMethod}

In this section, we present our method that can be used for a facial graph representation in an image collection. The method follows Fig. \ref{Fig:BlockDiagram}. Before we perform the analysis, it is necessary to do some preliminary computations. After that, we define some matrices that can be used to measure the `connectivity' between the subjects that are present in the collection, and finally we estimate and define the graph representation.

\subsection{Preliminary Computations: Feature Extraction} The idea of our approach is to analyze a set $\mathcal{I}$ of $n$ images $\{ {\bf I}_t \}$, for $t=1 \cdots n$. We detect all faces of $\mathcal{I}$ using Multi-task Cascaded Convolutional Networks (MTCNN) \cite{zhang2016joint} %\footnote{See implementation on \url{https://github.com/kpzhang93/MTCNN\_face\_detection\_alignment}.} 
that has been demonstrated to be very robust in unconstrained environments. % against poses, illuminations, expressions  and occlusions. 
All detected faces are stored as set $\mathcal{F}$ of $m$ face images $\{ {\bf F}_k \}$, for $k=1 \cdots m$. In addition, we store in vector ${\bf z}$ of $m$ elements the image index of the detected face image, \ie $z_k = t$, if face image ${\bf F}_k$ was detected in image ${\bf I}_t$.
Furthermore, the $m$ bounding boxes of the detected faces are stored in matrix ${\bf B}$ of $m$ \rx 4 elements with coordinates ${\bf b}_k = (x_1,y_1,x_2,y_2)_k$ for face image $k$. After face detection is performed, for each face image $k$, we compute the width and height ($w_k,h_k$) of the bounding box. 
Afterwards, we compute: \ri  the age ($a_k$), the gender ($g_k$) %Many models based on convolutional neural networks have been trained in the last years with promising results. 
using the library py-agender \cite{Py-Agender} 
%\footnote{\url{https://pypi.org/project/py-agender/}.} 
that offers very good results. % for age and gender estimation. The age, given in years, is estimated as a real number, and the gender is estimated as a real number between 0 and 1 (greater than 0.5 means female, otherwise male). 
\rii The facial expressions (${\bf e}_k$) are defined as a vector of seven probabilities \cite{arriaga2017real} for: angry, disgust, scared, happy, sad, surprised, and neutral. %Thus, ${\bf e}_k$ can be defined as the 7-element vector of expression for face $k$.
\riii The 68 facial landmarks (${\bf l}_k$) give the coordinates $(x,y)$ of the eyebrows (left and right), eyes (left and right), nose, mouth and jawline. For this end, we use the library Dlib \cite{king2009dlib}, %\footnote{See example \url{http://dlib.net/face\_landmark\_detection.py.html}} with very good results. 
\riv the pose vector (${\bf v}_k$) defines the direction where the face is looking to, for this end we use the projection of the roll vector obtained in \cite{Ruiz_2018_CVPR_Workshops}, %\footnote{See implementation \url{https://github.com/natanielruiz/deep-head-pose}.}, 
and \rv the face descriptor of $d$ elements (${\bf x}_k$), in this case we use descriptor with uni-norm.
%, \ie $||{\bf x}_k|| = 1$. In our experiments, we used many trained models (like VGG \cite{parkhi_2015}, FaceNet \cite{schroff_2015}, OpenFace \cite{amos2016openface}, Dlib \cite{king2009dlib} and ArcFace \cite{deng2019arcface}). Our conclusion is that 
In our experiments, we use ArcFace \cite{deng2019arcface}, that computes an embedding of $d=512$ elements with outstanding results. % comparing its performance to human vision in many complex scenarios. 
The descriptors of $m$ detected faces are stored in matrix ${\bf X}$  of $m \times 512$ elements.
In addition, we have a list of $n_e$ enrolled subjects that we want to analyze. It can be defined manually or using a face clustering algorithm. %The idea of face clustering is to build subsets (clusters) of faces that belong to the same identity. 
The descriptors of the enrolled faces have uni-norm and they are stored in a matrix ${\bf X}_e$ of $n_e \times 512$ elements. It is very simple to detect if the enrolled subjects are present in the image collection: we compute the $m \times n_e$-element matrix ${\bf Y} = {\bf X}{\bf X}_e^{\sf T} > \theta$, where $\theta$ is a threshold that we set to 0.4. If $Y(k,i) = 1$, that means that subject $S_i$ was detected in image $z_k$.

\subsection{Analysis: Connectivity Matrices} The following matrices have $n_e$ \rx $n_e$ elements. %, where $n_e$ is the number of subjects that will be analyzed. 
The indices of the elements of the matrices are $(i,j)$. Element $(i,j)$ gives a `connectivity measure' between subjects $S_i$ and $S_j$. The matrices are symmetric, that means for matrix ${\bf X}$, $X_{ij}=X_{ji}$. In addition, $X_{ii} = 0$. In all connectivity matrices of our approach, a high/low value of $X_{ij}$ means that the connectivity between $S_i$ and $S_j$ is high/low.

\noindent{\em 1. Co-occurrence Matrix} {\bf (C)}: Element $C_{ij}$ is defined as the number of images in the collection in which subjects $S_i$ and $S_j$ are present. This matrix is easily computed by defining a `presence matrix' ${\bf P}$ of $n_e \times n$ elements %(where $n$ is the number of images of the collection) 
in which element $P_{it}$ is 1/0 if subject $S_i$ is present/absent in image ${\bf I}_t$.

\noindent{\em 2. Closeness Matrix} {\bf (D)}: Element $D_{ij}$ is defined as the sum of the `closeness factors' of subjects $S_i$ and $S_j$ in those images of the collection in which $S_i$ and $S_j$ are present. The factor is computed as follows: For subjects $S_i$ and $S_j$, we measure the size in pixels of the bounding box of the faces as $a_i = \sqrt{w_ih_i}$ and $a_j = \sqrt{w_jh_j}$, the average $(a_i+a_j)/2$, and the distance $d$ in pixels between the centers of the bounding boxes in the image. The `closeness factor' is computed for those images that have similar face sizes ($\min(a_i/a_j,a_j/a_i) < 0.7$) and for the faces that are close enough ($d /a < n_f$). In our experiments, we set $n_f$ to 4. With the first criterium, we can avoid faces that are close in the image but because of the perspective they are far away in 3D space. With the second criterium, we consider only those pairs of images that are close enough. In our case, the faces must be closer than $4a$ (4 `faces'). Thus, the `closeness factor' is defined as $(n_f-d/a)/n_f$.  In this case, the factor is close to one, if the faces are very close; it is for example 0.25, if the distance of the faces is $3a$, and it is zero if the distance is greater than $4a$.

\noindent{\em 3. Connection Matrix} {\bf (Z)}: Element $Z_{ij}$ is defined as the sum of the `connection factors' of subjects $S_i$ and $S_j$ in those images of the collection in which $S_i$ and $S_j$ are present. The factor is computed as follows: For subject $S_i$ and $S_j$, we compute the intersection of vectors ${\bf v_i}$ and ${\bf v_j}$ defined as the projected roll vectors (computed from the head pose) that start at the point that is in the middle of both eyes (computed as the center of the landmarks corresponding to the eyes of the face). If the intersection of the vectors are in front of the faces, we compute both distances from intersection point to the middle of both eyes, and select the minimum ($d$). If $d$ is very small, it means that one subject is seeing close to the other one. The `connection factor' is defined (like the `closeness factor') as $(n_f-d/a)/n_f$ for $d/a<n_f$.

\noindent{\em 4. Empathy Matrix} {\bf (E)}: Element $E_{ij}$ is defined as the sum of the `empathy factors' of subjects $S_i$ and $S_j$ in those images of the collection in which $S_i$ and $S_j$ are present. The factor is computed as follows: For subject $S_i$ and $S_j$, we measure the first six elements of the expression vector (we avoid the neutral expression) as vectors ${\bf e}'_i$ and ${\bf e}'_j$. The idea of the `empathy factor' is to be close to one (or zero), if the expressions of subjects $S_i$ and $S_j$ are similar (or different). For this end, we use the cosine similarity, that means, we normalize the vectors as ${\bf e}'_i/||{\bf e}'_i||$ and ${\bf e}'_j/||{\bf e}'_j||$ and define the `empathy factor' as the dot product of them.

\noindent{\em 5. Happiness Matrix} {\bf (H)}: Element $H_{ij}$ is defined as the sum of the `happiness factors' of subjects $S_i$ and $S_j$ in those images of the collection in which $S_i$ and $S_j$ are present. The factor is computed as follows: For subjects $S_i$ and $S_j$, we extract the fourth element of the expression vector (that corresponds to the happiness) $h_i = e_i(4)$ and $h_j = e_j(4)$, and define the `happiness factor' as the average $(h_i+h_j)/2$.

Finally, the total connectivity matrix ${\bf T}$ is defined as the average: ${\bf T} = ({\bf C}+{\bf D}+{\bf Z}+{\bf E}+ {\bf H})/5$. We tested different weighted sums with similar results.

%\begin{equation}
%% {\bf T} = \alpha_1 {\bf C} + \alpha_2 {\bf D} + \alpha_3 {\bf Z} + \alpha_4 {\bf E} + \alpha_5 {\bf H}
%{\bf T} = \frac{1}{5}({\bf C} +{\bf D} + {\bf Z} +  {\bf E} + {\bf H}).
%\label{Eq:ConnectivityMatrix}  
%\end{equation}
% \noindent where $\sum \alpha_i = 1$. In our experiments, we set all weight factors to 1/5.

\subsection{Graph Construction} In order to build a graph that represent the connectivity of the subjects to be analyzed, we define a graph of $n_e$ nodes (one for each subject) and locate them in a 2D space in position $(x_i,y_i)$ for $i=1 \cdots n_e$. The key idea of the graph, is that the closer are the nodes $i$ and $j$, the higher the connectivity between them. That means, the distance of both nodes $\Delta_{ij} = ||(x_i,y_i)-(x_j,y_j)||$ should be a value that represents the `no-connectivity' between subjects $S_i$ and $S_j$. In our approach, we use the `connectivity' defined in previous section (see matrix  ${\bf T}$). We tested several definitions for `no-connectivity' and the best one --in terms of the visualization of the graph-- was given by $W_{ij} = 1/\sqrt{Q_{ij}+1}$, where $Q_{ij} = 99 \times T_{ij}/\max{\bf (T)}$, that means the `no-connectivity' $W_{ij}$ for subjects $S_i$ and $S_j$ gives values between 0.1 (when the connectivity is maximal) and 1.0 (when there is no connectivity). In addition, we set $W(i,i) = 0$.
The idea of our method is to find the coordinates of the nodes $(x_i,y_i)$, so that the distance of the nodes $\Delta_{ij}$, should be similar to the `no-connectivity' $W_{ij}$. This is an optimization problem that can be represented by minimizing the Frobenius norm: error $ = ||{\bf \Delta} - {\bf W}||^2_F \rightarrow \min$. We solve this problem using the simplex search method \cite{fminsearch} with a starting point given by the solution of a graph drawing by force-directed placement \cite{forcegraph}, in which the length of each edge in the graph is proportional to its weight $W_{ij}$. In our experiments, we report the Mean Absolute Error (MAE), as a metric of the performance of the graph, defined as the mean difference $|\Delta_{ij}-W_{ij}|$ in all pairs of nodes. Obviously, MAE cannot be zero in many cases where (\eg $A$ is friend of $B$ and $C$, but $B$ and $C$ are enemies).

%for a large number of nodes, because for example the connectivity of subjects $(A,B)$ and $(A,C)$ can be very high (\ie the nodes $A,B,C$ should be very close), however, the connectivity of subjects $(B,C)$ can be very low.

% In previous section, we obtained the location of the nodes $(x_l,y_l)$ for $l=1 \cdots n_e$ for the $n_e$ subjects of our dataset. Now, we have the following alternatives for the graph representation. 
\begin{figure*}[t!]
\centering 
\includegraphics[height=8.2cm]{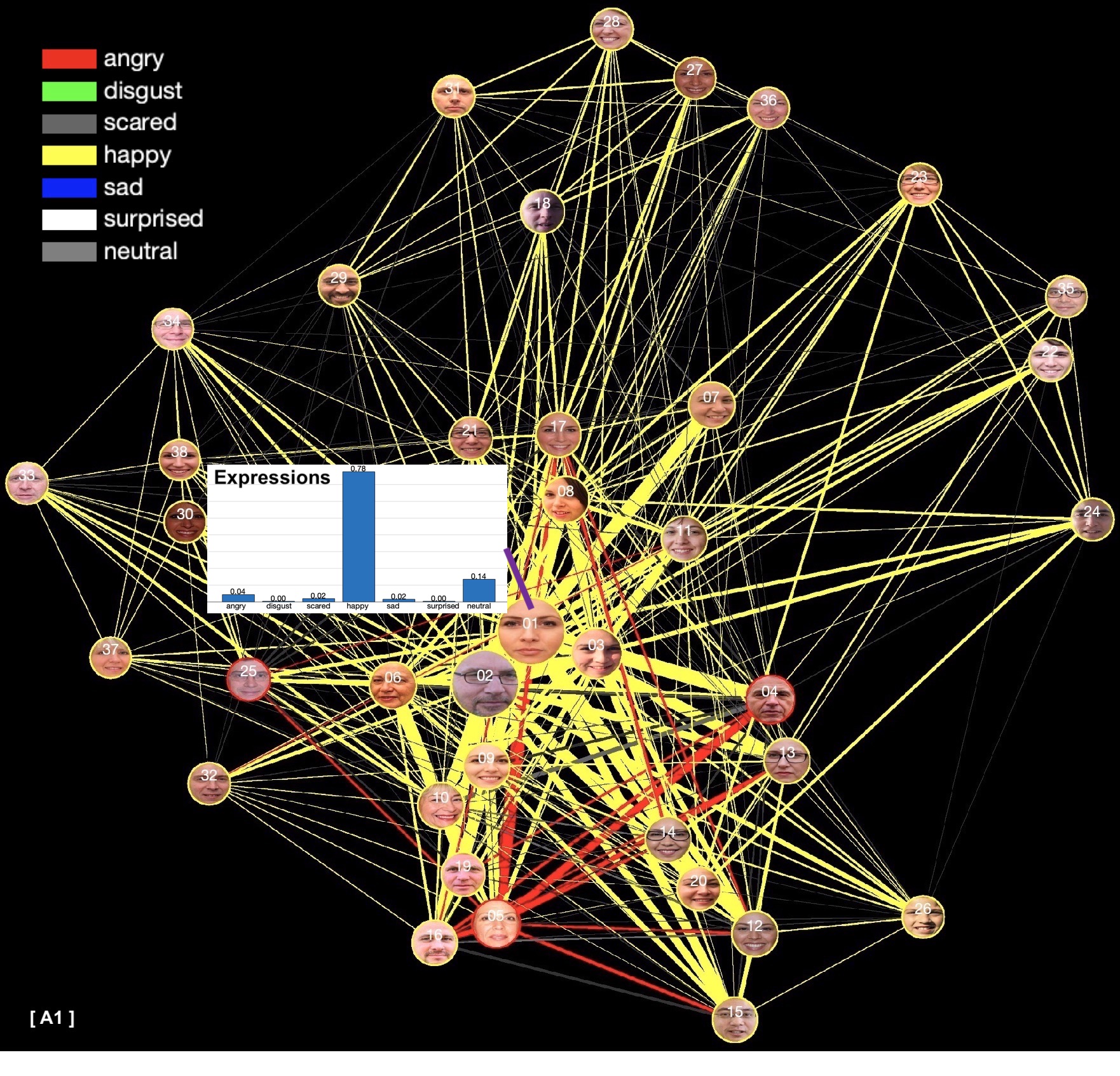}
\includegraphics[height=8.2cm]{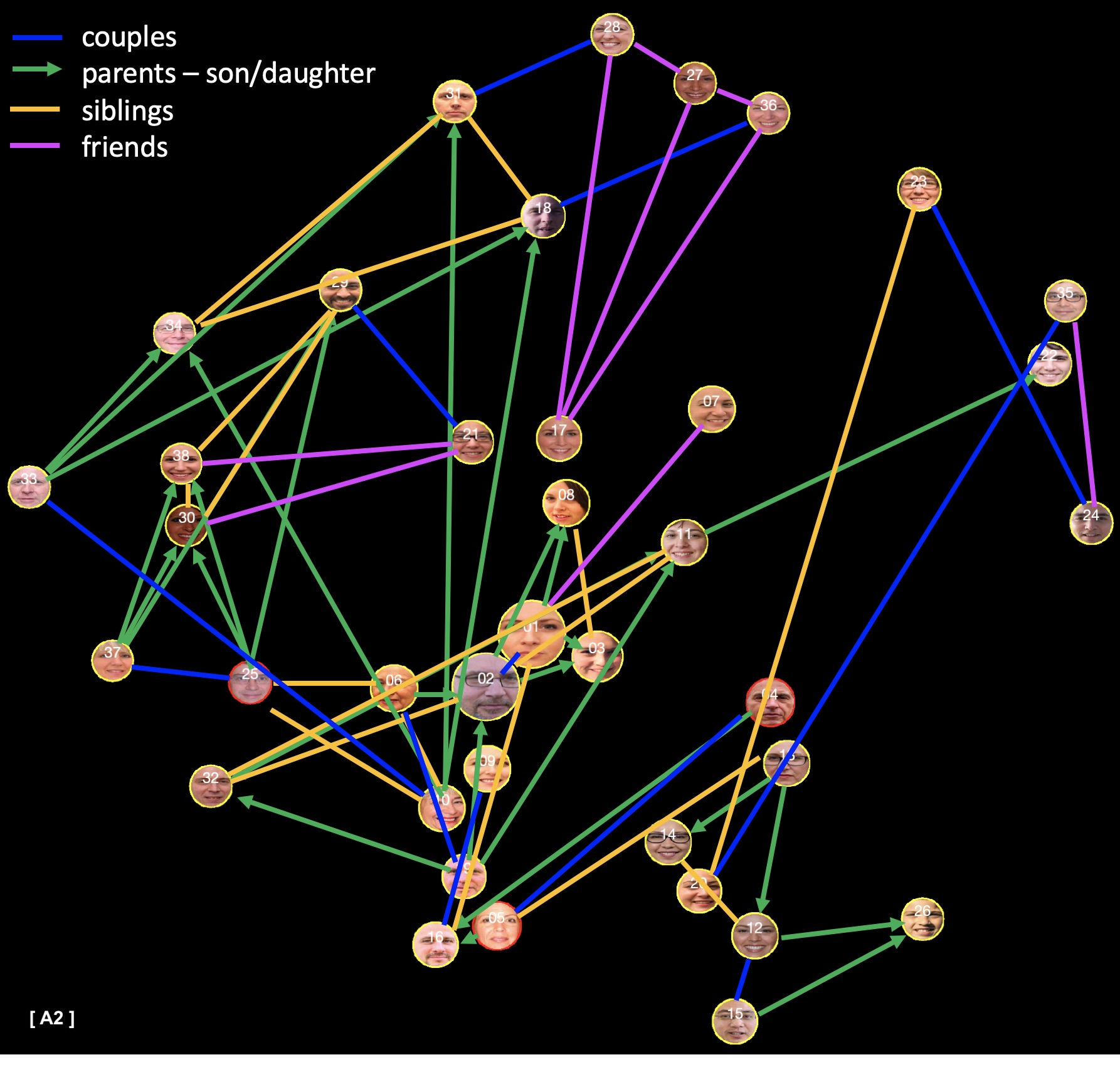}
\includegraphics[height=4.8cm]{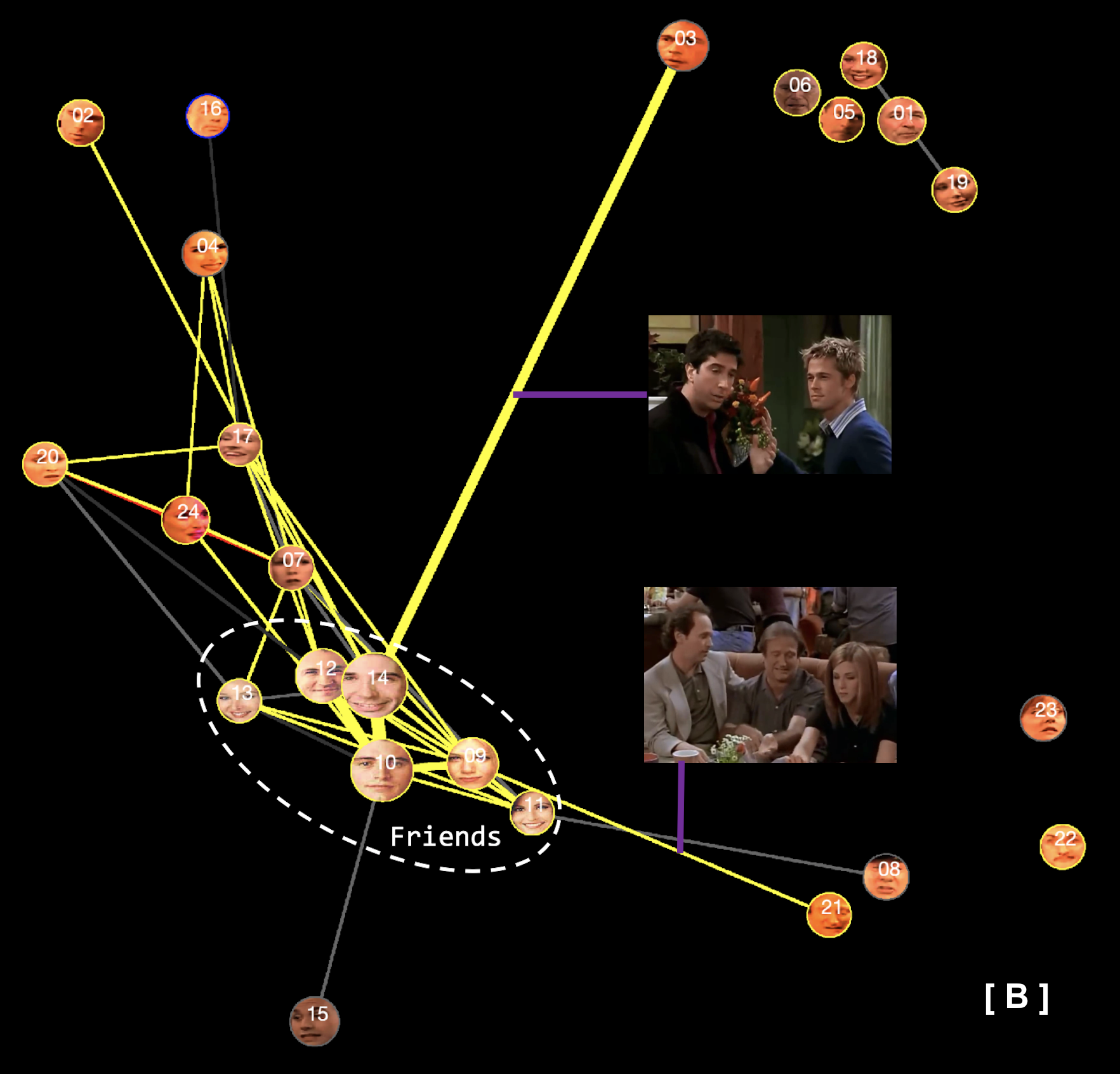}
\includegraphics[height=4.8cm]{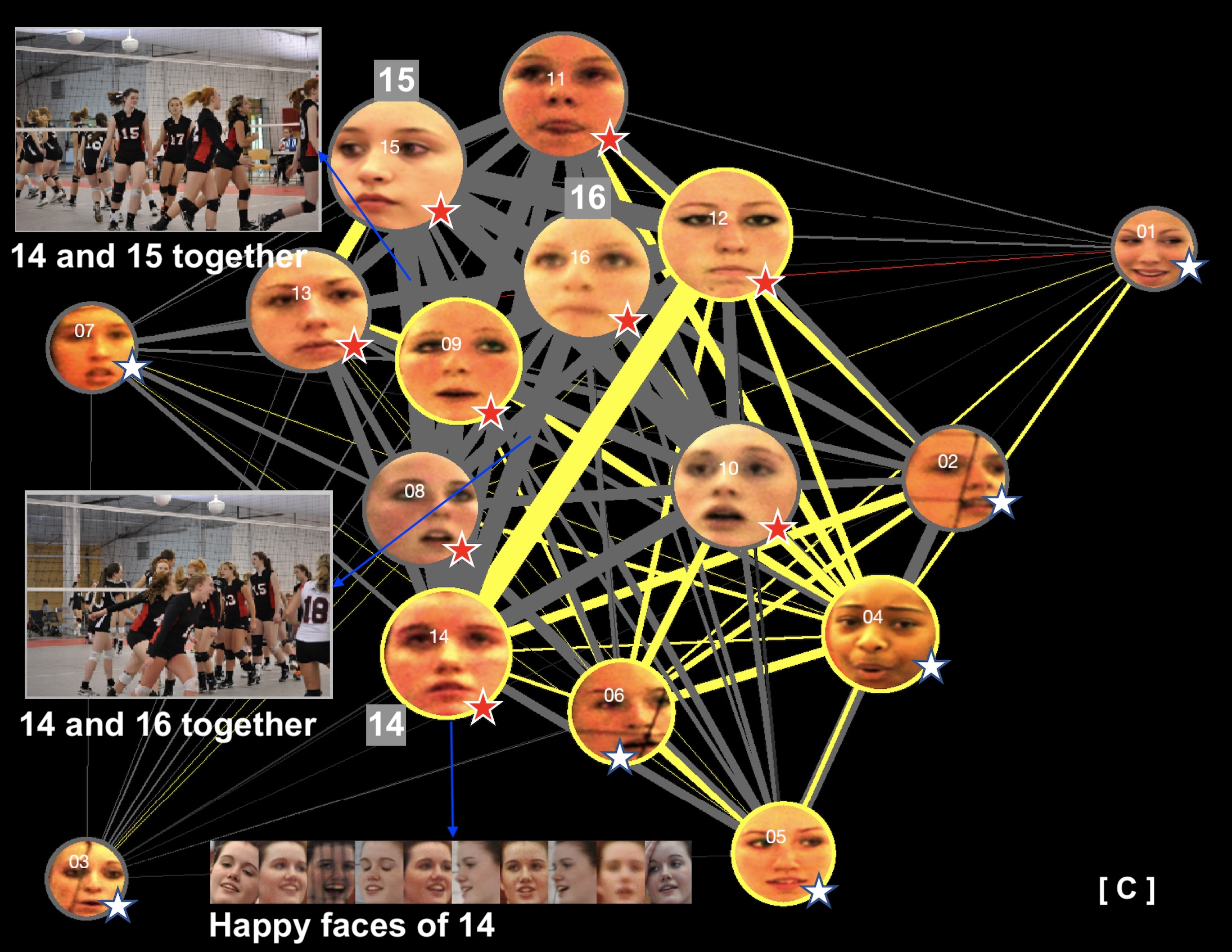}
\includegraphics[height=4.8cm]{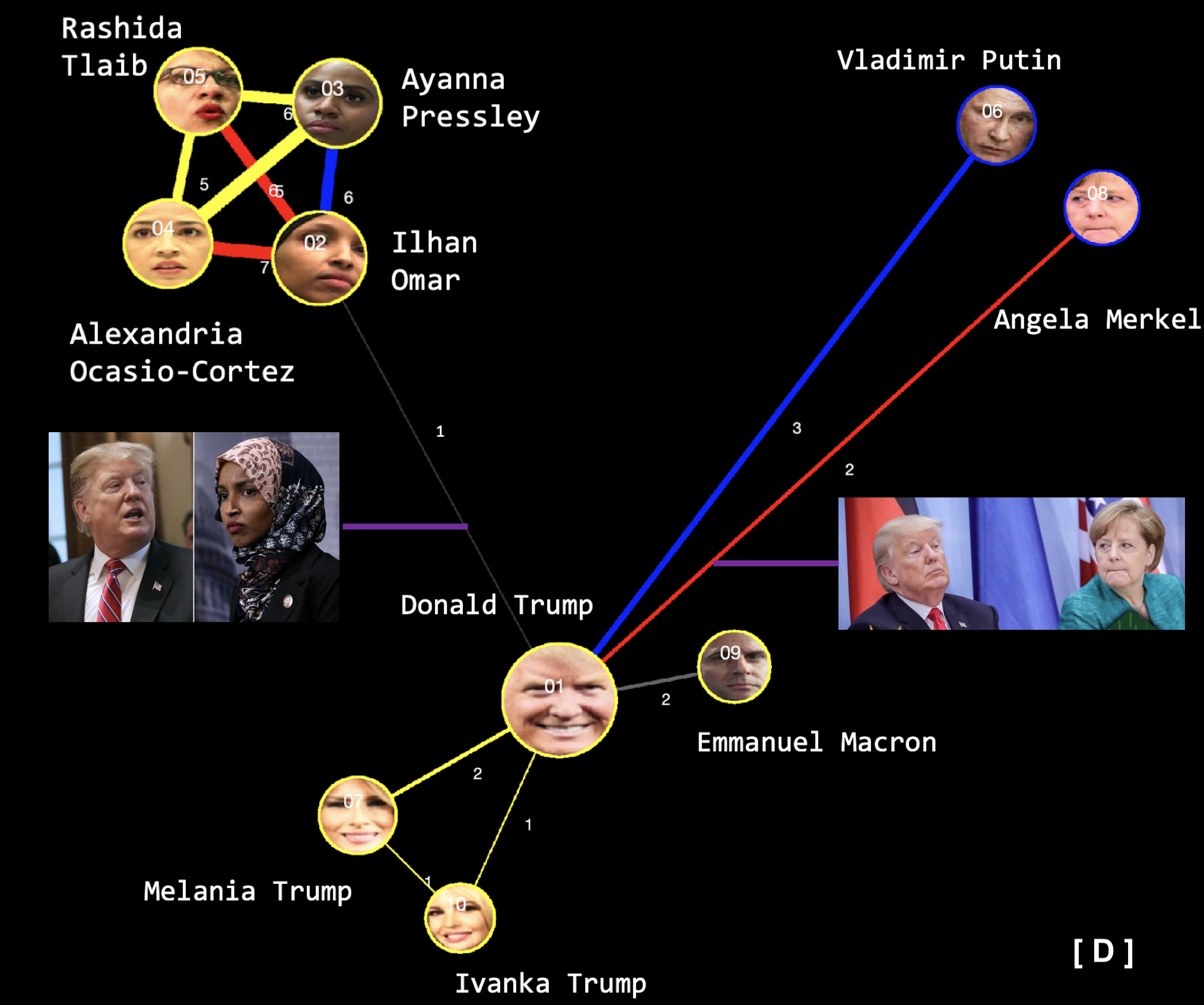}
\caption{Graph Representations. A) Wedding -- A1: proposed graph representation. A2: Nodes of proposed graph with edges given manually (blue lines: couples, green arrows: parents to sons and daughters, orange lines: siblings, magenta lines: friends. B) Friends (ellipse is plotted manually), C) Volleyball: two teams (red and white stars drawn manually). D) Trump - The Squad. In addition, some examples of the GUI (see text).}
\label{Fig:Graph}
\end{figure*}

\subsection {Graph Representation} For the node representation: a node is typically represented as a point or a circle. Alternatives in this case can be the size of the circle, color of the circle and color of the boundary of the circle. In our case for node $i$, we put the face of subject $S_i$ in a circle (the radius is related to the number of images in which $S_i$ is present), and the color of the boundary is related to the predominant facial expression of $S_i$. We use the colors based on  \cite{emotioncolor}: scare (dark gray), angry (red), disgusted (green), sad (blue), surprised (white), happy (yellow), neutral (gray).
For the edge representation: an edge is typically represented as a line, or (bi)directional arrow. Alternatives in this case can be the color and the width of the line. In our case for edge $(i,j)$, we decided to use lines (and not arrows) to make the graph simpler. In addition, our connectivity matrix is symmetric and it has no sense to use arrows\footnote{Nevertheless, the definition of connection given by matrix ${\bf Z}$ can be changed, if we consider the case when subject $S_i$ is looking to subject $S_j$ but not vice-versa. In this case, directional arrows can be used.}. The color of the lines, in our case, is set to the most common expression between subjects $S_i$ and $S_j$, and the width of the line corresponds to the number of images in which both subjects are present. If $S_i$ and $S_j$ do not have pictures together ($C_{ij} = 0$), there will be no edge between them.

Since the information of the co-occurrence matrix and presence matrix is stored, we propose to display the graph in a Graphical User Interface (GUI) in which the user can interact with the nodes and edges. By clicking onto a node, the user can obtain information related to the subject of the node: \eg gender, age and facial expressions. In addition, the images of the collection in which the subject is present can be displayed according to the facial expression (the images can be sorted using the expression `happiness' for example). On the other hand, by clicking onto an edge (that connect two nodes, in our case two subjects), the user can display the corresponding images of the connected subjects (images of the collection in which both subjects are present).

\section{Experimental results}
\label{Sec:Experiments}

% for socialanalysis.m 
% ss = 50; enroll = 11; % Friends3 (best)
% ss = 10; enroll = 19; % Volley (best)
% ss = 12; enroll = 6; % Matrimonio (best): real
% ss = 12; enroll = 17; % Matrimonio (best): fake
% ss = 42; enroll = 15; % Trump

In this section, we report the results obtained in four experiments that we used to validate the proposed approach. The implementation was done in Python 3.6.4 (for face detection and feature extraction) and Matlab R2019b (for connectivity matrices, graph construction and representation and Graphical User Interface).%\footnote{Code available at www... -- not given because of blind submission.}. %In each experiment, we report the mean average error, obtained as the main distance between the location obtained after the optimization (error = $||{\bf \Delta} - {\bf W}||^2_F$).

\subsection{Private Wedding} We took a private family album of a wedding celebration in which we participated. That means, we know exactly the subjects that are present and the relations that they have. In order to protect the privacy of the participants, we public only the attributes and descriptors of the faces, and for the graph representation we use synthetic faces generated by a GAN model \cite{karras2019style}.
%\footnote{We select manually faces that are similar to the originals from \url{https://www.thispersondoesnotexist.com}.}. 
This dataset has 639 images and 2280 faces. The obtained graph is illustrated in Fig. \ref{Fig:Graph}-A1. In Fig. \ref{Fig:Graph}-A2, we provide an additional graph with the same nodes, in which the edges are drawn manually according to the existing relation between the participants (couples, parents, sons, daughters, siblings and friends). 

\noindent $\rightarrow$ {\em Discussion:} (MAE = 0.1944) In the graph, we observe that main participants are the bride and the groom. In the GUI, we show the histogram of the expressions of the bride (she was happy in 78\% of the pictures). The main facial expression of the wedding is `happiness' (yellow), however, there are some people that is angry (the parents of the bride). It is very impressive, how close are the nodes in the graph for many of the strongest relationships of the participants.

\subsection{Friends} We downloaded videos of the sitcom `Friends' with a compilation of the six main characters (Rachel, Phoebe, Monica, Joey Chandler and Ross) and guest stars (like Julia Roberts, Brad Pitt, Robin Williams, etc.)\footnote{See both videos on \url{https://youtu.be/7GbOUIFa87g} and \url{https://youtu.be/8mP5xOg7ijs}.}. This dataset has 16429 images and 27647 faces. For the enrollment, we select the six main characters and 18 guest stars. The obtained graph is illustrated in Fig. \ref{Fig:Graph}-B. 

\noindent $\rightarrow$ {\em Discussion:} (MAE = 0.2313) From the graph, we observe that the main participants coincide with the six main characters of Friends. The main facial expression in the graph for this comedy is `happiness' (yellow). There is for example a connection with Brad Bitt (he was present in 363 images, and from them, Ross was present in 67 as well). In the GUI, we can display the images with co-occurrences.

\subsection{Volleyball Game} On July 2nd, 2019, we downloaded from Flickr the album ``VBVBA RVC 2 2010'' of pictures taken by Bruce A Stockwell \footnote{See the album \url{https://www.flickr.com/photos/bas68/albums/72157624234584197}. The pictures are licensed under a Creative Commons ``Attribution-NonCommercial-NoDerivs 2.0 Generic''}. In these pictures, we observe pictures of different volleyball games played on April 2010 by teenage players.  This dataset has 1131 images and 4550 faces. The obtained graph is illustrated in Fig. \ref{Fig:Graph}-C. In this set, there are two teams of players (see red and white stars included manually in the graph). For the enrollment, we select the players of both teams. 

\subsection{Discussion:} (MAE = 0.0679). 16\% and 6\% of the detected faces belong to the red and white teams respectively. We observe in the graph, that both teams are clustered, and the main participants (size of the faces) are from the red team. The facial expressions are `scared' (gray) and `happiness' (yellow). This is very typical in a game that stress and happiness co-exist. The GUI shows images of some of the main participants. In addition, we show the images of player 14 sorted according to her `happiness'.  

\subsection{Donald Trump -- The Squad} On July 19th, 2019, we downloaded images from Twitter given the hashtags {\tt \#Trump} and {\tt \#DonaldTrump}. In those days, there was a problem between Trump and four Democratic congresswomen (known as {\em The Squad}). This dataset has 494 images (126 were automatically removed because they were duplicated), 
and 677 faces. The obtained graph is illustrated in Fig. \ref{Fig:Graph}-D. For the enrollment we select Trump and some relatives, some politicians and the four congresswomen. 

\noindent $\rightarrow$ {\em Discussion:} (MAE = 0.1251). In the graph, we observe that main participant is Trump, he has a `happy' relation to his wife and daughter, and a `sad' and `angry' relation to Putin and Merkel respectively. An independent cluster is given by {\em The Squad} with only one connection to Trump (given by a photo montage). For this connection we use dark gray to show that there is no common facial expression.

 \section{Conclusions}

In this paper, we presented a graphical tool that can be used to detect and analyze social relations in an image collection. We proposed an optimal graph representation that is based on the `connectivity' of the subjects. We based our measurement on co-occurrence, closeness, facial expressions, the orientation of the head. The nodes represent the subjects of the collection, and the edges are their connectivities. In our solution, the closer the nodes, the more connected are the subjects. We propose a representation for the nodes and edges (colors are related to facial expressions and size are related to presence). The graph can be used in a graphical user interface (GUI) in which we can display the original images that shows the connection of the people. Finally, we present relevant results by analyzing a wedding celebration, a sitcom video, a volleyball game and images extracted from twitter given a hashtag. We believe that this tool can be very helpful to detect the existing social relations in an image collection. For future work, we would like to expand to three or more people by analyzing the connectivity matrix.

\ifCLASSOPTIONcompsoc
  \section*{Acknowledgments}
\else
  \section*{Acknowledgment}
\fi
The authors want to thanks to Grant Fondecyt--Chile \# 1191131.

\ifCLASSOPTIONcaptionsoff
  \newpage
\fi

\bibliographystyle{IEEEtran}

\end{document}